\documentclass{ws-procs961x669}
\usepackage{amssymb}
\usepackage{amsmath}
\usepackage{verbatim}
\usepackage{enumerate}
\usepackage{color}
\usepackage{graphicx}
\usepackage{xcolor}
\usepackage{amscd}

\newcommand{\gh}{\mathrm{gh}}
\newcommand{\T}{\mathbb{T}}
\newcommand{\Z}{\mathbb{Z}}
\newcommand{\CC}{\mathbb{C}}
\newcommand{\F}{\mathcal{F}}
\newcommand{\M}{\mathcal{M}}
\newcommand{\HH}{\mathcal{H}}
\newcommand{\LLL}{\mathsf{L}}
\newcommand{\LL}{\mathcal{L}}
\newcommand{\RR}{\mathfrak{R}}
\newcommand{\R}{\mathbb{R}}
\newcommand{\B}{\mathcal{B}}
\newcommand{\res}{\mathrm{res}}
\newcommand{\HDens}{\mathrm{Dens}^{\frac12}}
\newcommand{\ra}{\rightarrow}
\newcommand{\xra}{\xrightarrow}
\newcommand{\pa}{\partial}
\newcommand{\bt}{\bullet}
\newcommand{\mr}{\mathrm}
\newcommand{\din}{\partial_\mathrm{in}}
\newcommand{\dout}{\partial_\mathrm{out}}
\newcommand{\bA}{\mathbb{A}}
\newcommand{\bB}{\mathbb{B}}

\begin{document}
\title{
Perturbative BV theories with Segal-like gluing
}
\author{Alberto~S.~Cattaneo}
\address{Institut f\"ur Mathematik, Universit\"at Z\"urich,
Winterthurerstrasse 190, CH-8057 Z\"urich, Switzerland\\
e-mail: alberto.cattaneo@math.uzh.ch}

\author{Pavel Mnev}
\address{
Max Planck Institute for Mathematics, Vivatsgasse 7,
53111 Bonn, Germany;\\
St. Petersburg Department of V. A. Steklov Institute of Mathematics of\\ the Russian Academy of Sciences, Fontanka 27, St. Petersburg, 191023 Russia\\
e-mail: pmnev@pdmi.ras.ru}

\author{Nicolai Reshetikhin}
\address{Department of Mathematics,
University of California, Berkeley,
CA 94720, USA;\\
ITMO University. Saint Petersburg 197101, Russia; \\
KdV Institute for Mathematics, University of Amsterdam,\\ 1098 XH Amsterdam, The
Netherlands\\ 
e-mail: reshetik@math.berkeley.edu}

\begin{abstract}
This is a survey of our 
program of perturbative quantization of gauge theories on manifolds with boundary compatible with cutting/pasting and with gauge symmetry treated by means of a cohomological resolution (Batalin-Vilkovisky) formalism. We also give two explicit quantum examples -- abelian $BF$ theory and the Poisson sigma model. 
This exposition is based on a talk by P.M. at the ICMP 2015 in Santiago de Chile.
\end{abstract}
\bodymatter
\section{Introduction: calculating partition functions by cutting-pasting}
Locality in quantum field theory can be understood as the possibility to recover, by means of a simple gluing formula, the partition function on a manifold $M$ split into submanifolds $M_1,\ldots, M_k$ from partition functions on the pieces $M_1,\ldots,M_k$. This principle was made precise in the setting of 2-dimensional conformal field theory by Segal\cite{Segal} and for topological field theory by Atiyah\cite{Atiyah}. In this description, an $(n-1)$-manifold $\Sigma$ gets assigned a complex vector space -- the space of states $\HH_\Sigma$, and an $n$-manifold $M$ 
with boundary $\pa M$ gets assigned a vector $Z_M\in \HH_{\pa M}$ in the space of states for its boundary. The main axiom states that if $M=M_1\cup_\Sigma M_2$ is the gluing of two $n$-manifolds along a closed $(n-1)$-manifold $\Sigma$, the partition function for the whole manifold can be recovered from the partition functions for the pieces, $Z_M=\langle Z_{M_1},Z_{M_2}\rangle_\Sigma$. Here $\langle,\rangle_\Sigma$ is the pairing of states in $\HH_\Sigma$. In the case of a topological theory, partition functions are interesting diffeomorphism invariants of manifolds that can be recovered from cutting the (possibly  complicated) manifold into simple pieces. For instance, for a 2-dimensional topological field theory, it suffices to know the space of states for the circle and the partition function for the disk and the pair of pants to recover the partition function on any oriented surface.

Our program\cite{1dCS,CMR1,CMR4,CMR5} is to construct quantum field theories on manifolds with boundary, compatible with cutting-pasting in Atiyah-Segal sense, from perturbative path integral quantization of gauge theories  
in the Batalin-Vilkovisky formalism. 

Our next goals are 1) to extend this quantization procedure to cutting-pasting with corners and 2) to prove that the
  $k\to\infty$ asymptotics of the Reshetikhin-Turaev invariant\cite{RT} coincides with the perturbative expansion of Chern-Simons theory\cite{Witten
  }.

\section{BV-BFV formalism for gauge theory on manifolds with boundary: an outline}
\subsection{Classical BV-BFV formalism}
A classical $n$-dimensional BV-BFV theory \cite{CMR1} is defined, in the spirit of Atiyah-Segal axiomatics of QFT \cite{Atiyah,Segal}, as the following association $\mathbb{T}$. 
\begin{itemize}
\item 
To a closed $(n-1)$-manifold $\Sigma$, the theory $\T$ assigns a \textit{phase space} $\Phi_\Sigma$ -- a supermanifold equipped with 
\begin{itemize}
\item $\Z$-grading by the ghost number,\footnote{In bosonic theories, the parity 
is the mod 2 reduction of the $\Z$-grading. The $\Z$-grading is useful for bookkeeping, but is not really essential, and is not even available in some field theories\cite{1dCS}.
}
\item a cohomological vector field $Q_\Sigma$ (an odd vector field of ghost number $\gh(Q_\Sigma)=1$ satisfying $(Q_\Sigma)^2=0$) -- the \textit{BRST operator},\footnote{Geometrically, it is a vector field; it also an operator in the sense that it acts on functions on the phase space.}
\item an even exact symplectic structure $\omega_\Sigma$ of ghost number $0$, compatible with $Q_\Sigma$, with a fixed primitive $1$-form $\alpha_\Sigma$ such that $\omega_\Sigma=\delta \alpha_\Sigma$,\footnote{We use $\delta$ to denote de Rham operator on fields and reserve $d$ for de Rham on the spacetime manifold. 
In a more general setup, $\omega_\Sigma$, rather than being exact, is allowed to be the curvature of a connection in a $U(1)$-bundle over the phase space.
}
\item the \textit{BFV charge} $S_\Sigma$ -- an odd function of ghost number $1$ which is the Hamiltonian for $Q_\Sigma$.
\end{itemize}
\item To an $n$-manifold $M$ with possibly nonempty boundary, $\T$ assigns the \textit{space of fields} $\F_M$ -- a $\Z$-graded supermanifold equipped with the following structures.
\begin{itemize}
\item Boundary restriction of fields -- a projection (surjective submersion) $\pi: \F_M\ra \Phi_{\pa M}$,
\item a cohomological vector field $Q_M$; it is required to be projectable by $\pi$, with the boundary phase space BRST operator its projection, $\pi_* Q_M=Q_{\pa M}$,
\item an odd symplectic structure $\omega_M$ of ghost number $-1$ (the \textit{BV 2-form}),
\item the \textit{action} (or \textit{master action}) $S$ -- an even function of ghost number $0$, satisfying the following ``almost-Hamiltonianity'' relation:
\begin{equation}\label{main eq}
\iota_{Q_M}\omega_M=\delta S_M+\pi^*\alpha_{\pa M}
\end{equation}
\end{itemize}
\item Disjoint unions of manifolds are mapped by $\T$ to direct products of phase spaces/spaces of fields.
\item If an $n$-manifold $M$ is cut along a codimension $1$ submanifold $\Sigma$ into two pieces $M_1$ and $M_2$, then the space of fields on the whole $n$-manifold is the (homotopy) fiber product of spaces of fields for pieces $M_{1}$ and $M_2$ over the phase space for the cut $\Phi_\Sigma$, \quad  $\F_M=\F_{M_1}\times_{\Phi_\Sigma} \F_{M_2}$.
\end{itemize}

The main structure equation (\ref{main eq}) implies that the BRST operator $Q_M$ does not preserve the BV 2-form $\omega_M$, instead the Lie derivative is a boundary term: $L_{Q_M}\omega_M=\pi^* \omega_{\pa M}$. Another consequence of (\ref{main eq}) is a form of Batalin-Vilkovisky classical master equation:
$\frac12\, \iota_{Q_M}\iota_{Q_M}\omega_M=\pi^* S_{\pa M}$.

\begin{remark} One can pass to the ``reduced'' BV-BFV picture, by passing to the  \textit{Euler-Lagrange moduli spaces} $\M_M$, $\M_\Sigma$ -- generally, singular super varieties, constructed as the zero locus of $Q$ quotiented by the integrable distribution induced by $Q$ on the zero-locus.\footnote{E.g. in abelian Chern-Simons theory (with gauge group $\mathbb{R}$) on a 3-manifold with boundary, the relevant moduli spaces are given by de Rham cohomology with degree shift, $\M_M=H^\bt(M)[1]$, $\M_\Sigma=H^\bt(\Sigma)[1]$. For non-abelian Chern-Simons, they get replaced by certain natural super-geometric extension of the moduli space of flat connections on $M$ and $\Sigma$, respectively.} 
Under some Hodge-theoretic assumptions on the BV-BFV theory, $\M_\Sigma$ carries an even-symplectic structure $\underline{\omega}_\Sigma$, the image of $\pi_*: \M_M\ra \M_\Sigma$ is Lagrangian, $\M_M$ carries a $\gh=1$ Poisson structure whose symplectic leaves are fibers of $\pi_*$, $\M_\Sigma$ carries a prequantum $U(1)$-bundle $\LLL_\Sigma$ with connection $\nabla_\Sigma$ (inherited from $\alpha_\Sigma$) of curvature $\underline{\omega}_\Sigma$ and the pullback bundle $(\pi_*)^*\LLL_\Sigma$ over $\M_M$ carries a horizontal section (understood as the exponential of the \textit{Hamilton-Jacobi action}).
\end{remark}


\subsection{Quantum BV-BFV formalism}\label{sec: qBV-BFV}
A quantum $n$-dimensional BV-BFV theory\cite{CMR4} is the following association $\T_q$.
\begin{itemize}
\item To a closed $(n-1)$-manifold, $\T_q$ assigns a \textit{BFV space of states} -- a cochain complex of $\CC$-vector spaces $\HH^\bt_\Sigma$ graded by the ghost number, with differential $\Omega_\Sigma$ (the \textit{quantum BFV charge}).
\item To an $n$-manifold $M$ with boundary, $\T_q$ assigns:
\begin{itemize}
\item a finite-dimensional \textit{space of residual fields}\footnote{Cf. ``slow'' (or ``infrared'')  fields in Wilson's effective action approach to renormalization. Also, in our examples, ``residual fields'' are the same as ``zero-modes''.} $\F_M^\res$ equipped with a BV 2-form (an odd $\gh=-1$ symplectic structure) $\omega_M^\res$,
\item the \textit{partition function} -- an element in the space of states for the boundary valued in half-densities of residual fields
$Z_M\in \HDens(\F_M^\res)\otimes \HH_{\pa M}$
satisfying the \textit{BV quantum master equation} (QME), modified by a boundary term:
\begin{equation}\label{mQME}
\left(\Omega_{\pa M}+\hbar^2\Delta_M^\res\right)Z_M=0
\end{equation}
where $\Delta_M^\res$ is the \textit{canonical BV operator} -- the second order odd Laplacian on half-densities on $\F_M^\res$ associated to the odd symplectic structure $\omega_M^\res$. Operator acting on $Z_M$ in (\ref{mQME}) is required to square to zero. The partition function $Z_M$ is defined modulo equivalence 
\begin{equation}\label{gauge equiv}
Z_M\sim Z_M+\left(\Omega_{\pa M}+\hbar^2\Delta_M^\res\right)(\cdots)
\end{equation}
coming from the gauge-fixing ambiguity.
\end{itemize}
\item Disjoint unions are sent by $\T_q$ to tensor products (for spaces of states and partition functions) and direct products (for residual fields).
\item If $M$ is cut into pieces $M_1$ and $M_2$ by a codimension $1$ submanifold $\Sigma$, then the partition function on $M$ is recovered by the following 
procedure:
\begin{enumerate}[(i)]
\item One constructs $\widetilde{Z}_M=\langle Z_{M_1}, Z_{M_2}\rangle_\Sigma$ where $\langle,\rangle_\Sigma$ denotes the pairing of states in $\HH_\Sigma$.\footnote{More precisely, 
it is the canonical pairing between the space of states $\HH_\Sigma$ and its dual, as $\Sigma$ embeds into $M_1$ and $M_2$ with opposite orientations. Reversal of orientation of a $(d-1)$-manifold acts on the space of states by dualization.}
\item $\widetilde{Z}_M$ is a half-density on $\F_{M_1}^\res\times \F_{M_2}^\res$ (with values in vectors in $\HH_{\pa M}$). To obtain a half-density on a smaller space $\F_M^\res$, one splits $\F_{M_1}^\res\times \F_{M_2}^\res$ into $\F_M^\res$ and a symplectic complement $W$ and evaluates the integral over a Lagrangian $\LL$ in $W$,
$Z_M=\int_\LL \widetilde{Z}_M$. We call this fiber BV integral construction the \textit{BV pushforward}\cite{CMR4} $P_*$ of half-densities along the odd symplectic fibration $P:\F_{M_1}^\res\times \F_{M_2}^\res\ra \F_M^\res$. Thus, the final gluing formula is
\begin{equation}\label{gluing formula}
Z_M=P_*\,\langle Z_{M_1}, Z_{M_2}\rangle_\Sigma
\end{equation}
\end{enumerate}
\end{itemize}

\begin{remark}
A correction to this picture is that one may allow different \textit{realizations} of the space of residual fields $\F_M^\res$, taking values in the \textit{
partially ordered set (poset)
of realizations} $\RR_M$. Then if $r_1\succ r_2$ is an ordered pair of realizations, one can pass from $r_1$ to $r_2$ by a BV pushforward $Z^{r_2}_M=P_* Z^{r_1}_M$ corresponding to an odd symplectic fibration of a bigger model for residual fields over the smaller one $P: \F_M^{r_1}\ra \F_M^{r_2}$. Jumping along the poset of realizations by BV pushforwards is a model for Wilson's renormalization group flow (in that context, realizations correspond for values of momentum cutoff). In special examples\cite{CMR5}, one can construct realizations corresponding to cellular decompositions of a manifold, with poset structure given by cellular aggregations (inverses of subdivisions).
\end{remark}

\subsection{Quantization -- the idea}
The general idea of the passage from a classical BV-BFV theory $\T$ to a quantum one $\T_q$ is as follows.  Here we assume for simplicity that the spaces of fields are graded vector spaces (as opposed to more general graded manifolds).\footnote{This assumption makes perfect sense in perturbation theory, where the perturbative path integral sees only a formal neighborhood  of a fixed classical solution of equations of motion.}

For an $(n-1)$-dimensional closed manifold $\Sigma$, one fixes a fibration of the phase space $p:\Phi_\Sigma\ra \B_\Sigma$ with Lagrangian fibers. Moreover, one requires that the primitive 1-form $\alpha_\Sigma$ vanishes on fibers of $p$. Then one defines the space of states as the space of $\CC$-valued half-densities on the base $\HH_\Sigma^\bt=\HDens_\CC(\B_\Sigma)$. This is a simple instance of geometric quantization. 
The differential on $\HH_\Sigma^\bt$ (the  quantum BFV charge) is constructed as a quantization of the classical BFV charge $\Omega_\Sigma=\widehat{S}_\Sigma$. In many examples there is a \textit{preferred} quantization, defined as a series in $\hbar$, which does square to zero and gives the correct boundary term for the QME (\ref{mQME}).

For an $n$-manifold $M$ with boundary, we consider fibers $\F_b\subset \F_M$ of the composition $\F_M\xra{\pi} \Phi_{\pa M}\xra{p} \B_{\pa M}$  over 
$b\in \B_{\pa M}$, i.e. $\F_b$ are fields on $M$ with boundary values in the Lagrangian fiber $p^{-1}\{b\}\subset \Phi_{\pa M}$. It is tempting to define the partition function as a function of the boundary condition $b$, by a functional integral $Z_M(b)=\int_{\LL\subset \F_b}e^{\frac{i}{\hbar}S_M}\mu_M^{\frac12}$ over a gauge-fixing Lagrangian $\LL\subset \F_b$; here $\mu_M^{\frac12}$ is a reference half-density on $\F_M$. However, such an integral is typically perturbatively ill-defined due to zero-modes of the quadratic part of $S_M$. The solution is to split out a finite-dimensional subspace $\F_M^\res$ out of $\F_b$, i.e. fix a splitting $\F_b=\F_M^\res\times W$ compatible with the BV 2-form, and integrate over a Lagrangian $\LL$ in $W$: 
$$Z_M(b,\phi)=\int_{\LL\subset W}e^{\frac{i}{\hbar}S_M}\mu_M^{\frac12}$$
Here $\phi\in \F_M^\res$ is a residual field. The result is a complex half-density on $\B_{\pa M}$ and a half-density on $\F_M^\res$:
$$Z_M\in \CC\otimes \HDens(\B_{\pa M})\otimes \HDens(\F_M^\res)=\HH_{\pa M}\otimes \HDens(\F_M^\res)$$
In a class of examples\cite{CMR4}, one can prove that the perturbative (Feynman diagram) evaluation of $Z_M$ satisfies the axioms of a quantum BV-BFV theory of Section \ref{sec: qBV-BFV} (QME, cohomological independence on the choice of gauge-fixing, gluing formula).

\section{
Some topological examples
}
\subsection{Abelian $BF$ theory}\label{sec: abBF}
In abelian $BF$ theory\cite{Schwarz} on an $n$-manifold $M$, fields are pairs of differential forms $\F_M=\Omega^\bt(M)[1]\oplus \Omega^\bt(M)[n-2]\;\ni (A,B)$; the BV 2-form pairs the two summands, $\omega_M=\int_M \delta B\wedge \delta A$. The action is $S_M=\int_M B\wedge dA$ and the cohomological vector field is $Q_M=\int_M dA\wedge \frac{\delta}{\delta A}+dB\wedge \frac{\delta}{\delta B}$. For $M$ with boundary split into in- and out-part, $\pa M=\din M\sqcup \dout M$ (a \textit{cobordism}), we correct the action by a boundary term to $S_M=\int_M B\wedge dA+(-1)^{n-1}\int_{\din} B\wedge A$. The phase space $\Phi_{\pa M}\ni (A_\pa,B_\pa)$ is the space of pairs of forms on $\pa M$ and the base of Lagrangian fibration is $\B_{\pa M}=\Omega^\bt(\dout M)[1]\oplus \Omega^\bt(\din M)[n-2]\ni (\bA,\bB)$.

The space of states of the theory is $\HH_{\pa M}=\HDens_\CC(\B_{\pa M})$. In particular, it contains states of the form 
$$\!\!\!\!\!\!\!\!\!\! 
\int_{C_j(\din M)\times C_k(\dout M)\ni (x_1,\ldots,x_j;y_1,\ldots,y_k)} \Psi(x_1,\ldots,x_j;y_1,\ldots,y_k)\cdot \bB(x_1)\cdots \bB(x_j)\cdot \bA(y_1)\cdots \bA(y_k)$$
where $C_j,C_k$ are the configuration spaces of $j$ distinct points $x_1,\ldots,x_j$ on in-boundary and $k$ distinct points $y_1,\ldots,y_k$ on out-boundary; $\Psi$ are the coefficient functions (``wave-functions'') which parameterize the states. More generally one can allow sums of such expressions for different $j,k$ and insertions of 
monomials 
in $\bA,\bB$ at points of the boundary, rather than fields $\bA,\bB$ themselves.

The quantum BFV operator on the space of states is simply the lifting of the de Rham operator $\Omega_{\pa M}=(-1)^{n}i\hbar\left(\int_{\din M}d\bB\wedge \frac{\delta}{\delta \bB}+\int_{\dout M}d\bA\wedge \frac{\delta}{\delta \bA}\right)$.

The space of residual fields is the double of de Rham cohomology relative to the boundary components $\F_M^\res=H^\bt(M,\dout M)[1]\oplus H^\bt(M,\din M)[n-2]\;\ni (a,b)$. 
It inherits an odd symplectic form given by Poincar\'e-Lefschetz duality. 
Explicit calculation of the partition function yields\cite{CMR4}:
\begin{equation*}\label{Z abBF}
\!\!\!\!\!\!\!\!\!\!\!
Z_M=\xi_M\cdot\tau(M,\dout M)\cdot e^{\frac{i}{\hbar}\left((-1)^{n-1}\int_{\din M}\bB\wedge a+(-1)^n\int_{\dout M} b\wedge \bA-\int_{\din M\times \dout M\ni (x,y)} \bB(x)\wedge \eta(x,y)\wedge \bA(y)\right)}
\end{equation*}
Here $\eta\in \Omega^{n-1}(C_2(M))$ is the propagator -- the integral kernel of the homotopy inverse of de Rham operator on forms on $M$ vanishing on $\dout M$; $\tau(M,\dout M)\in \mr{Det}\,H^\bt(M,\dout M)/\{\pm 1\}$ is the Reidemeister torsion of $M$ relative to $\dout M$. 
Note that the determinant line $\mr{Det}\,H^\bt(M,\dout M)/\{\pm 1\}$ is canonically identified with constant half-densities on $\F_M^\res$. 
The coefficient\cite{CMR5}
$$
\xi_M = (2\pi\hbar)^{\sum_{k=0}^n(-\frac14-\frac12 k(-1)^k)\cdot\dim H^k(M,\dout M)}\cdot (e^{-\frac{\pi i}{2}}\hbar)^{\sum_{k=0}^n(\frac14-\frac12 k(-1)^k)\cdot\dim H^k(M,\dout M)} 
$$
contains a mod 16 phase $e^{\frac{2\pi i}{16}s}$ with $s=\sum_{k=0}^n(-1+2k(-1)^k)\dim H^k(M,\dout M)$, which bears some similarity with the Atiyah-Patodi-Singer eta invariant appearing in the phase of Chern-Simons partition function\cite{Witten}.
The partition function $Z_M$ 
satisfies the QME (\ref{mQME}), changes by an equivalence (\ref{gauge equiv}) with the change of gauge-fixing (choice of propagator $\eta$ and choice of representatives for cohomology) and behaves with respect to cutting/pasting according to the gluing formula (\ref{gluing formula}).

\subsection{The Poisson sigma model}
Let $\pi=\sum_{\alpha,\beta=0}^m \pi^{\alpha\beta}(u)\frac{\pa}{\pa u^\alpha}\wedge \frac{\pa}{\pa u^\beta}$ be a Poisson bivector field on $\mathbb{R}^m$. 
The Poisson sigma model\cite{Strobl,Ikeda,Kontsevich,CF} is a 2-dimensional sigma model defined by the BV action
$$S(A,B)=\int_M \sum_\alpha B^\alpha\wedge dA_\alpha +\sum_{\alpha,\beta}\frac12 \pi^{\alpha\beta}(B)\wedge A_\alpha\wedge A_\beta 
$$
where the fields $(A_\alpha,B^\beta)\in (\Omega^\bt(M)[1]\oplus \Omega^\bt(M)[n-2])\otimes \mathbb{R}^m$ are the $m$-component versions of the fields of abelian $BF$ theory. Thus, the Poisson sigma model is a perturbation of the ($m$-component) 2-dimensional abelian $BF$ theory by an interaction term depending on a Poisson bivector field on the target $\R^m$.

For a surface with boundary $\pa M=\din M\sqcup \dout M$, the space of states $\HH_{\pa M}$ is the same as for abelian $BF$ theory (where the fields $\bA_\alpha, \bB^\alpha$ now carry the target index).
The residual fields are the $m$-component version of those of Section \ref{sec: abBF}, $\F_M^\res=(H^\bt(M,\dout M)[1]\oplus H^\bt(M,\dout M)[n-2])\otimes \R^m\;\ni (a_\alpha, b^\alpha)$.

The partition function is as follows:
\begin{equation*}
Z_M=\xi_M^m\cdot \tau(M,\dout M)^m\cdot \exp\frac{i}{\hbar}\left(\sum_\Gamma \frac{(-i\hbar)^{\mr{loops}(\Gamma)}}{|\mr{Aut}(\Gamma)|}\int_{C_{j,k,l}(M)}\phi_\Gamma(\bA,\bB,a,b)\right)
\end{equation*}
Here $\xi_M$ and $\tau(M,\dout M)$ are the same as in Section \ref{sec: abBF}. The sum in the exponential is over oriented connected graphs $\Gamma$ without short loops\footnote{This is consistent with the assumption that either $\pi$ is unimodular or the surface 
has zero Euler characteristic.} with $j\geq 0$ 1-valent vertices on $\din M$ with adjacent half-edge oriented \textit{from} the vertex, $k\geq 0$ 1-valent vertices on $\dout M$ with adjacent half-edge oriented \textit{to} the vertex, $l\geq 0$ internal vertices on $M$ with 2 outgoing and $\geq 0$ incoming half-edges. The graph is allowed to have loose half-edges (leaves). Half-edges are decorated with target space index $\alpha$; in-vertices -- with $\bB^\alpha$, out-vertices -- with $\bA_\alpha$, bulk vertices of valence $(2,r)$ -- with partial derivatives of $\pi$ at the origin, $\left.\frac{\pa^r}{\pa u^{\beta_1}\cdots \pa u^{\beta_r}}\right|_{u=0}\pi^{\alpha_1\alpha_2}(u)$. Edges are decorated with the propagator $-\delta^\alpha_\beta\cdot\eta(x,y)$, with $\eta$ as in Section \ref{sec: abBF}. 
Leaves -- with residual fields $a_\alpha$ (for out-orientation), $b^\alpha$ (for in-orientation). Wedging the forms associated with vertices, edges and 
leaves, 
one obtains a differential form $\phi_\Gamma(\bA,\bB,a,b)$ on the compactified configuration space $C_{j,k,l}(M)$ of $j+k+l$ distinct ordered points on $M$ such that $j$ of them are on $\din M$ and $k$ of them are on $\dout M$. Form $\phi_\Gamma$ is polynomial in boundary fields $\bA_\alpha,\bB^\alpha$ and residual fields $a_\alpha,b^\alpha$ and the integral over the configuration space is convergent.
\begin{figure}[h!]
\begin{center}\includegraphics[scale=0.46]{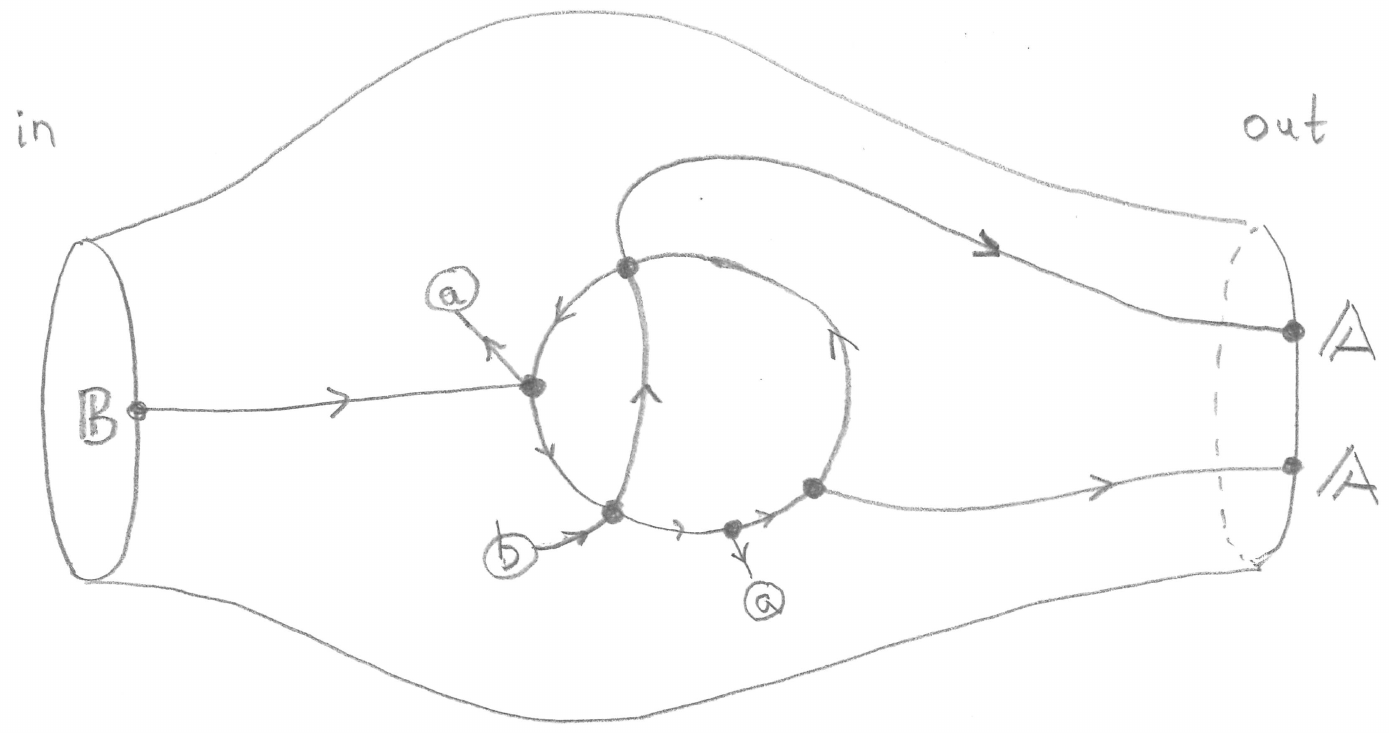}\end{center}
\caption{A typical graph $\Gamma$ contributing to $Z_M$.}
\end{figure}

The differential $\Omega_{\pa M}$ 
on $\HH_{\pa M}$ can be calculated from the boundary contributions of configuration space integrals appearing in the partition function: acting on $Z_M$, $\Omega_{\pa M}$ is the standard-ordering quantization (replacing $\bA_\alpha\mapsto i\hbar \frac{\delta}{\delta \bB^\alpha}$ on $\din M$ and $\bB^\alpha\mapsto i\hbar \frac{\delta}{\delta \bA_\alpha}$ on $\dout M$, and putting all derivatives to the right) of the expression
$$\int_{\pa M} \sum_\alpha\bB^\alpha\wedge d\bA^\alpha+\sum_{\alpha,\beta}\frac12 \Pi^{\alpha\beta}(\bB)\wedge \bA_\alpha\wedge \bA_\beta$$
where $\Pi^{\alpha\beta}(u)=\frac{u^\alpha*u^\beta-u^\beta*u^\alpha}{i\hbar}$ is the deformation of $\pi$ by Kontsevich's star-product\cite{Kontsevich}.

These data ($Z_M$, $\Omega_{\pa M}$) satisfy the properties of a quantum BV-BFV theory -- the QME (\ref{mQME}), cohomological 
independence on gauge-fixing (\ref{gauge equiv}) and gluing (\ref{gluing formula}).

\section*{Acknowledgements}
A. S. C. acknowledges partial support of SNF Grant No. 200020-149150/1. This research was (partly) supported by the NCCR SwissMAP, funded by the Swiss National Science Foundation, and by the COST Action MP1405 QSPACE, supported by COST (European Cooperation in Science and Technology). P. M. acknowledges partial support of RFBR Grant No. 13-01-12405-ofi-m. Research of N. R. was partially supported by the NSF Grant DMS- 0901431 and by RFBR Grant No. 14-11-00598.

\end{document}